\documentclass[aps,pra,reprint,superscriptaddress,amsmath]{revtex4-1}

\usepackage[utf8x]{inputenc}
\usepackage{graphicx}
\usepackage{dcolumn}
\usepackage{amsmath}
\usepackage{xcolor}
\usepackage{physics}

\begin{document}


\title{Atom interferometry using $\sigma^+ -\sigma^-$ Raman transitions between \newline $\ket{F=1,m_F=\mp1}$ and $\ket{F=2,m_F=\pm1}$}

\author{J. Bernard}
\email{jeanne.bernard@onera.fr}
\affiliation{DPHY,ONERA, Universit\'{e} Paris Saclay, F-91123 Palaiseau, France}
\affiliation{LCM-CNAM, 61 rue du Landy, 93210, La Plaine Saint-Denis, France}


\author{Y. Bidel}
\affiliation{DPHY,ONERA, Universit\'{e} Paris Saclay, F-91123 Palaiseau, France}

\author{M. Cadoret}
\affiliation{LCM-CNAM, 61 rue du Landy, 93210, La Plaine Saint-Denis, France}

\author{C. Salducci}
\affiliation{DPHY,ONERA, Universit\'{e} Paris Saclay, F-91123 Palaiseau, France}

\author{N. Zahzam}
\affiliation{DPHY,ONERA, Universit\'{e} Paris Saclay, F-91123 Palaiseau, France}

\author{S. Schwartz}
\affiliation{DPHY,ONERA, Universit\'{e} Paris Saclay, F-91123 Palaiseau, France}

\author{A. Bonnin}
\affiliation{DPHY,ONERA, Universit\'{e} Paris Saclay, F-91123 Palaiseau, France}

\author{C. Blanchard}
\affiliation{DPHY,ONERA, Universit\'{e} Paris Saclay, F-91123 Palaiseau, France}

\author{A. Bresson}
\affiliation{DPHY,ONERA, Universit\'{e} Paris Saclay, F-91123 Palaiseau, France}


\date{\today}

\begin{abstract} 
We report on the experimental demonstration of a horizontal accelerometer based on atom interferometry using counterpropagative Raman transitions between the states $\ket{F=1,m_F=\mp1}$ and $\ket{F=2,m_F=\pm1}$ of $^{87}$Rb.  Compared to the $\ket{F=1, m_F=0} \leftrightarrow \ket{F=2,  m_F=0}$ transition usually used in atom interferometry, our scheme presents the advantages to have only a single counterpropagating transition allowed in a retroreflected geometry, to use the same polarization configuration than the magneto-optical trap and to allow the control of the atom trajectory with magnetic forces. We demonstrate horizontal acceleration measurement in a close-to-zero velocity regime using a single-diffraction Raman process with a short-term sensitivity of $25 \times 10^{-5}$ m.s$^{-2}$.Hz$^{-1/2}$. We discuss specific features of the technique such as spontaneous emission, light-shifts and effects of magnetic field inhomogeneities. We finally give possible applications of this technique in metrology or for cold-atom inertial sensors dedicated to onboard applications.
\end{abstract}


\maketitle




\section{Introduction}
 
During the last two decades, light-pulse atom interferometers (LPAIs) \cite{Kasevich1991} which exploit the wave-like interference of atoms, have become unique instruments for precision measurements of inertial forces, with applications in both applied and fundamental science. For example, atom interferometric techniques have been employed in measurements of gravitational  \cite{Fixler2007,Rosi2014} and fine structure constants \cite{Muller2018,Morel2020}, test of the equivalence principle \cite{Kasevich2020}, searches for dark sector particles \cite{Hamilton2015,Elder2016,Jaffe2018}, and even proposed for use in gravitational wave detection \cite{Dimopoulos2008,Canuel2018}. They have also enabled the realization of high performance accelerometers \cite{Canuel2006,Geiger2011,Xu2017,Bernard2019}, gyroscopes \cite{Gustavson1997,Durfee2006,Gauguet2009,Stockton2011,Tackmann2012,Berg2015,Dutta2016,Savoie2018}   and gravimeters \cite{Peters2001,Gillot2014,Hu2013,Hauth2013,Karcher2018} demonstrating great promise for fielded inertial sensors based on atom interferometry \cite{Bidel2018,Bidel2020,Wu2019}. In addition, they can also be utilized for probing the field gradient of an external field such as gravity \cite{McGuirk2002,Sorrentino2014,Duan2014} or magnetic fields \cite{Luo2011,Hardman2016}.

In a LPAI, sequences of laser pulses are used to split, deflect and recombine matter-waves to create atom interference. In inertial sensors, these sequences of light pulses commonly use counterpropagating two-photon Raman transitions with large one-photon detuning \cite{Kasevich1991} between hyperfine ground states of alkali atoms (e.g $\ket{F=1}$ and $\ket{F=2}$ for $^{87}$Rb). They form the basic atom-optics elements by finely controlling the external degree of freedom of the atoms through the generation of coherent superposition of momentum states. In a counterpropagating configuration, the transfer between the two internal ground states is always accompanied with a change of $\pm \hbar k_\textrm{eff}$ of the momentum state, where $k_\textrm{eff}$ is the effective wave vector.

In order to achieve high precision measurements, the two counterpropagating Raman lasers are usually obtained thanks to a retroreflected geometry where a single laser beam with two laser frequencies is retroreflected off a mirror. This geometry allows to mitigate parasitic effects induced by wave front distortions which are critical to achieve good accuracy and long term stability \cite{Peters2015,Karcher2018}. It also reduces interferometer phase noise as most vibration effects are common to the two laser fields. In addition, in order to avoid systematic errors induced by first order Zeeman effect, the atoms are commonly manipulated in the magnetically-insensitive  $m_F=0$ sublevels in the interferometer.

In this work, we report on the experimental realization of a Raman transition-based LPAI between magnetically-sensitive internal states in a Mach-Zehnder type geometry. Using the supplementary internal degree of freedom of atoms manipulated in sensitive magnetic sublevels, we realize a sensor which simultaneously measures inertial and magnetic accelerations. Our work focuses on the specific case of $^{87}$Rb. Using a $\sigma^+-\sigma^-$ polarized light arrangement, we manipulate the atoms in the interferometer  between the two magnetically-sensitive ground states $5S_{1/2}$ $\ket{F=1,m_F=\mp1} \to \ket{F=2, m_F=\pm1}$, also used as atomic clock transition for magnetically trapped $^{87}$Rb \cite{Cornell2002}, taking benefit of their similar first-order Zeeman shift. Using this technique we perform the measurement of the horizontal component of acceleration, in a close-to-zero velocity regime, using a single-diffraction Raman scheme \cite{Bernard2019}, without need for alternative techniques to lift the degeneracy of the double diffraction process \cite{Leveque2009}. We demonstrate a short-term sensitivity of $25 \times 10^{-5}$ m.s$^{-2}$.Hz$^{-1/2}$ for absolute acceleration measurement. We then discuss some specifics of our technique in comparison  with usual magnetically-insensitive Raman-based atom interferometers, such as spontaneous emission rate, additional sensitivity to magnetic field inhomogeneity and  light-shifts.
Finally, in light of the advantages of this technique, we propose atom interferometer designs which could be of interest in metrology, as well as for improving the performances of cold-atom inertial sensors in operational field conditions \cite{Bidel2018,Wu2019,Bidel2020,Geiger2011}.

\section{Method} \label{method}

\subsection{Principle and advantages of the method}

\begin{figure}[h]
\centering
\includegraphics[scale=0.55,trim=0cm 0cm 10cm 0cm,clip]{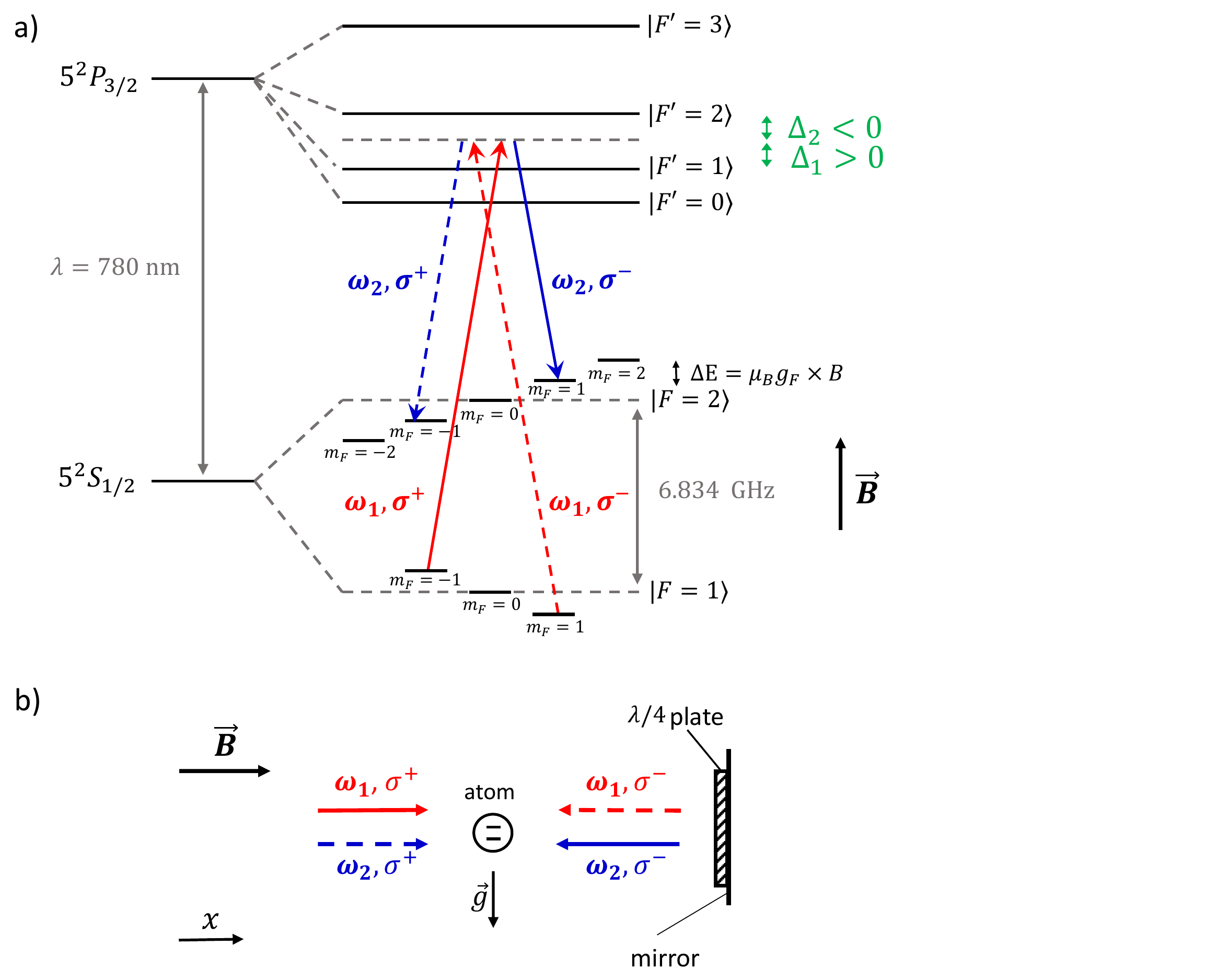}
\caption{(a) Scheme of the $\sigma^+ - \sigma^-$ Raman transitions between Zeeman sublevels of the two hyperfine ground states of $^{87}$Rb in the presence of a magnetic field. Solid lines represent the $\sigma^+ - \sigma^-$ polarized beams performing the $\ket{F=1,m_F=-1} \leftrightarrow \ket{F=2,m_F=1}$ transition. Dashed lines represent the $\sigma^- - \sigma^+$ polarized beams performing the $\ket{F=1,m_F=1} \leftrightarrow \ket{F=2,m_F=-1}$ transition. (b) Schematic setup of two-photon Raman transitions in the commonly used retroreflected geometry. A two-level atom is interacting with two pairs of counterpropagating light fields with $\sigma^+ - \sigma^-$ polarizations. This polarization arrangement allows for only  one pair of couterpropagating light fields to drive the Raman coupling  leading to a single diffraction Raman process despite zero Doppler shift.}
\label{figure1}
\end{figure}

We implement our method using a horizontal Mach-Zehnder type LPAI based on counterpropagative stimulated two-photon Raman transitions between the $\ket{F=1, m_F=\mp 1}$ and $\ket{F=2, m_F=\pm 1}$ hyperfine levels of the 5S$_{1/2}$ ground state of $^{87}$Rb. These two states are coupled via an intermediate state, using lasers of frequencies $\omega_{1(2)}$ detuned from the $\ket{5P_{3/2},F'=1(2)}$ state by the one-photon detuning $\Delta_{1(2)}$ (see Fig.\ref{figure1}(a)). A static bias magnetic field of magnitude $B$ aligned with the Raman lasers  is applied to define a quantization axis for the atoms. This field shifts the $\ket{F,m_F}$ ground state magnetic sublevels by $\Delta E=\mu_B g_F m_F B$ as a first approximation, where $F$ is the atomic total angular momentum, $m_F=0, \pm1,...,\pm F$ are its projections on the quantization axis, $\mu_B$ is the Bohr magneton, and $g_F$ is the Land\'e factor, equal to $-(+)1/2$ for the $F=1(2)$ states respectively. The two counterpropagating Raman beams are generated using a retroreflective setup. Contrary to many LPAI experiments using a lin$\perp$lin polarization configuration and a large one-photon detuning allowing to exclusively drive counterpropagative Raman transitions between the magnetic-insensitive $m_F=0$ sublevels, we implement here a $\sigma^+-\sigma^-$  configuration (see Fig.\ref{figure1}(b)): the Raman beams have a $\sigma^+$ polarization in one direction, and a $\sigma^-$ polarization in the retroreflected direction. The quantum state at the input of the interferometer is prepared to be in one single Zeeman sublevel $\ket{F=1,m_F}$ ($m_F=+1$ or $m_F=-1$). Thus, according to the electric dipole transition selection rules, only one counterpropagating transition is possible. Indeed the $\sigma^+ - \sigma^-$ Raman laser configuration only allows $\Delta m_F = \pm 2$ transitions. Consequently, the two-photon Raman transition couples the magnetically-sensitive hyperfine states $\ket{F=1,m_F=\pm1} \leftrightarrow \ket{F=2,m_F=\mp 1}$ with an effective Rabi frequency \cite{Steck}: 
\begin{equation} \label{eq1}
\Omega_\textrm{eff}= \Gamma^2 \frac{\sqrt{I_1 I_2}}{4 I_\textrm{sat}} \frac{1}{4 \sqrt{3}} \left( \frac{1}{\Delta_1} - \frac{1}{\Delta_2} \right)
\end{equation}

\noindent where $\Gamma=2\pi \times 6.07$ MHz is the natural line width, $I_\textrm{sat}=c \pi h \Gamma /3 \lambda^3=1.67$ mW.cm$^{-2}$ is the saturation intensity (with $c$ the speed of light, $h$ the Planck's constant and $\lambda=780$ nm), $I_{1(2)} $ are the Raman laser intensities and $\Delta_{1(2)}$ are the one-photon detunings with respect to the hyperfine levels $\ket{5P_{3/2},F'=1(2)}$ (see Fig.\ref{figure1}(a)). The Rabi frequency of the two-photon Raman transition constrains us to tune the one-photon transition in between $\ket{F'=1}$ and $\ket{F'=2}$ in order to avoid destructive interferences between the transition probability amplitudes for each excited states $\ket{F'=1}$ and $\ket{F'=2}$. Consequences of such a close-to-resonance detuning are discussed in Section \ref{ES}.

LPAIs usually manipulate atoms in the magnetically-insensitive $m_F=0$ sublevels. For zero-velocity atoms the use of retroreflected Raman beams leads naturally to a double-diffraction
scheme: two stimulated Raman transitions with opposite
momentum transfer $\pm \hbar \vec{k}_\textrm{eff}$ are simultaneously resonant \cite{Leveque2009}. Our scheme has the advantage of having only one counterpropagating Raman transition allowed despite the retroreflection. In addition, we can very easily implement the $k_\textrm{eff}$-reversal technique \cite{Peters2001} to eliminate some systematics by alternatively preparing the atoms in the $\ket{F=1,m_F=\mp1}$ states. Moreover the Raman beams have the same polarization as the magneto-optical trap beams, enabling a more compact and simple sensor.

\subsection{Experimental apparatus and lasers}

The experiment was carried out in the LPAI setup described in \cite{Perrin2019,Bernard2019}. The usual steps of atom interferometry (preparation, interferometry and population detection) were performed with the laser system described in \cite{Theron2017}. On the one hand an Erbium distributed feedback fiber laser (DFB-FL) at 1.5 $\mu$m, locked to a rubidium transition through a saturated absorption setup \cite{Theron2015}, is used to cool and detect the atoms. On the other hand a DFB laser diode at 1.5 $\mu$m, frequency controlled by a beat-note with the fiber laser, provides the LPAI laser source. The two Raman frequencies are generated using a fibered phase modulator \cite{Carraz2012}. Both lasers are finally combined at 1.5 $\mu$m through an electro-optical modulator acting like a continuous optical switch between each laser, before seeding a 5 W Erbium-doped fiber amplifier (EDFA). The output of the EDFA is sent to a second harmonic generation bench. The complete laser setup and optical bench description can be found in \cite{Bernard2019}.

\subsection{State preparation and Raman spectroscopy} \label{state preparation}

We investigate our method by selecting the atomic input state and implementing Raman spectroscopy. A cold $^{87}$Rb atom sample is produced in a three-dimensional magneto-optical trap (MOT) loaded from a background vapor in 840 ms. An optical molasses cools the atoms down to 2 $\mu$K in 8 ms. After turning off the cooling beams, the atoms are in free fall. Then a horizontal bias magnetic field $B \sim 400$ mG is switched on and a microwave $\pi$-pulse is applied, followed by a blow-away beam, allowing to select the atoms in the $\ket{F=1,m_F=-(+)1}$ state. Raman spectroscopy is performed using a Raman pulse of duration $\tau=10~ \mu$s. Finally an internal state-selective vertical light-induced fluorescence detection is used to measure the proportion of atoms in each hyperfine state $\ket{F=1}$ and $\ket{F=2}$. The cycling time of the experiment is $T_\textrm{cycle}=1$ s.

\begin{figure}[h!]
\centering
\includegraphics[scale=0.35]{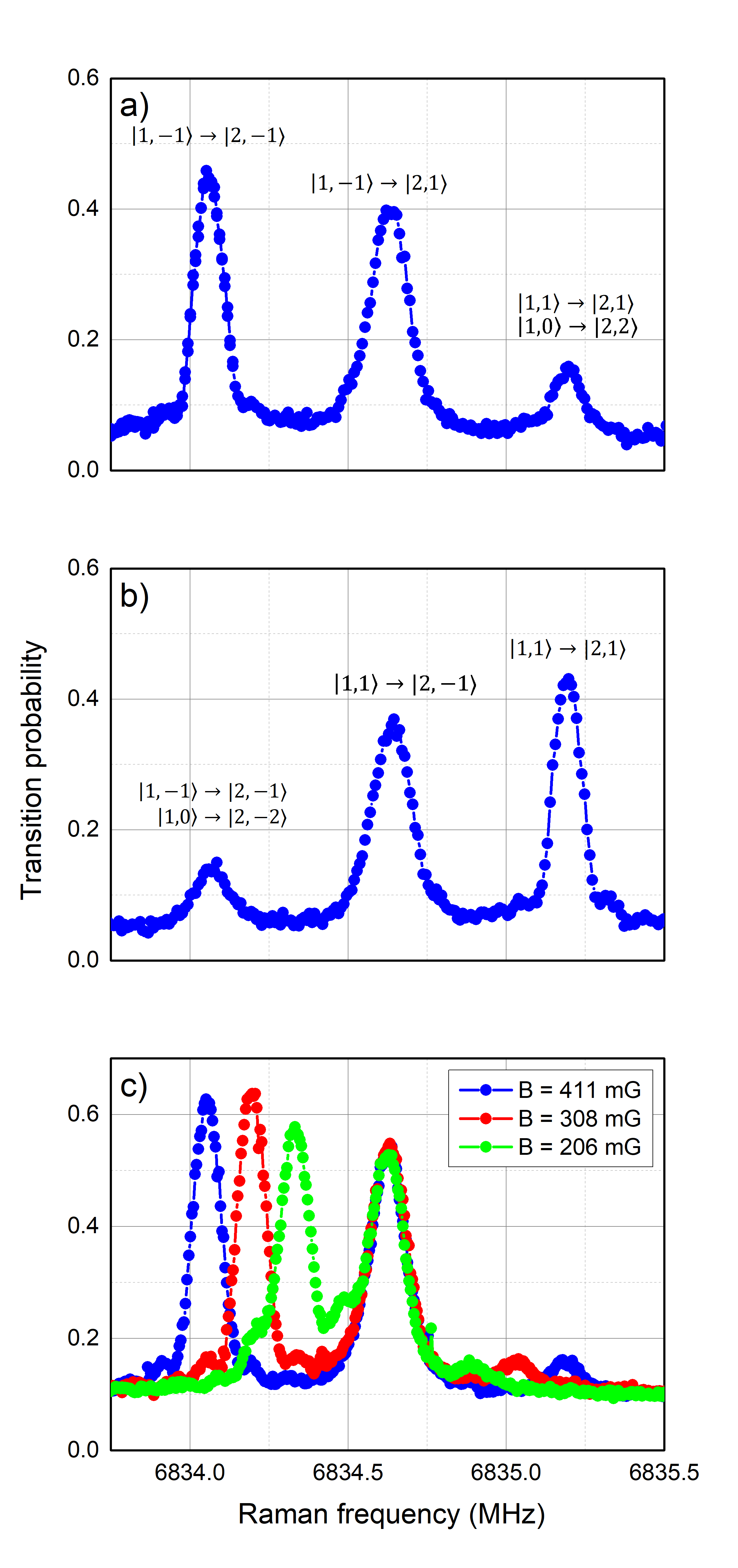}
\caption{Raman resonance spectra obtained by scanning the Raman frequency $(\omega_1-\omega_2)/2\pi$ across the resonances. (a) Raman spectrum with the atoms prepared in the $\ket{F=1,m_F=-1}$ sublevel. The magnetic field amplitude is $B=411$ mG. Three resonance peaks are observed and correspond to four transitions: $\ket{F=1,m_F=-1}\rightarrow\ket{F=2,m_F=-1}$, $\ket{F=1,m_F=-1}\rightarrow\ket{F=2,m_F=1}$ and two degenerate transitions $\ket{F=1,m_F=1}\rightarrow\ket{F=2,m_F=1}$ and $\ket{F=1,m_F=0}\rightarrow\ket{F=2,m_F=2}$. (b) Raman spectrum with the atoms prepared in the $\ket{F=1,m_F=+1}$ sublevel. The magnetic field amplitude is $B=411$ mG. (c) The atoms being prepared in the $\ket{F=1,m_F=-1}$ sublevel, the magnetic field amplitude is tuned from 200 mG to 400 mG. The transition $\ket{F=1,m_F=-1} \rightarrow \ket{F=2,m_F=1}$ is insensitive at first order to a magnetic field amplitude variation.}
\label{figure3}
\end{figure}

Figure \ref{figure3}(a) displays the measured transition probability as a function of the Raman frequency difference $(\omega_1-\omega_2)/2\pi$. The atoms being prepared in the $\ket{F=1,m_F=-1}$ state, the electric dipole transition selection rules state that only two transitions are possible with our Raman beam polarization configuration: a copropagating transition $\ket{F=1,m_F=-1}\rightarrow\ket{F=2,m_F=-1}$ (almost insensitive to Doppler effect and therefore narrower) and a counterpropagating transition $\ket{F=1,m_F=-1}\rightarrow\ket{F=2,m_F=1}$ (see Fig.\ref{figure1}(a)). A third transition can be observed in Fig.\ref{figure3}(a) due to spontaneous emission: a fraction of the atoms are transferred by spontaneous emission to the $m_F=0,1$ magnetic sublevels of $F=1$ and undergoes one of the two degenerate transitions $\ket{F=1,m_F=1}\rightarrow\ket{F=2,m_F=1}$, $\ket{F=1,m_F=0}\rightarrow\ket{F=2,m_F=2}$. Spontaneous emission will be further discussed in Section \ref{ES}. 

One can notice not only  that the $\sigma^+-\sigma^-$ transition frequency is independent of the magnetic field magnitude $B$ at first order (see Fig.\ref{figure3}(c)), but also that the $\pm k_\textrm{eff}$ transition frequencies (corresponding to $\ket{F=1,m_F=\mp1}\leftrightarrow\ket{F=2,m_F=\pm1}$) are the same (see Fig.\ref{figure3}(b)), which is very useful when implementing the $k_\textrm{eff}$-reversal technique (see Section \ref{fringes}). 

In conclusion, once the Raman frequency is properly tuned, it is only the state preparation in the magnetic sublevel $m_F=\mp1$ that defines which transition $\pm k_\textrm{eff}$ will be addressed during the interferometer. This means that reversing the effective wave vector is different from LPAIs using the $\ket{F=1,m_F=0}-\ket{F=2,m_F=0}$ transition, where the Raman frequency needs to be changed in order to reverse $ k_\textrm{eff}$.

\section{Atom interferometry with $\sigma^+ - \sigma^-$ transitions} \label{AI technique}

Our Mach-Zehnder type LPAI in a horizontal configuration consists of a $\pi/2-\pi-\pi/2$ Raman pulse sequence, with each pulse separated by a time $T$. Due to free fall of atoms across the laser beams of waist 5.5 mm ($1/e^2$ radius), the interrogation time is limited to $2T=33$ ms. At the output of the interferometer the phase shift is the sum of two terms: $\Delta \phi = \Delta \phi_\textrm{propagation} + \Delta \phi_\textrm{laser}$ where $\Delta \phi_\textrm{propagation}$ is the difference in the action computed along the classical trajectory of each interferometer arm, and $\Delta \phi_\textrm{laser}$ is the phase difference imprinted on the atoms by the Raman lasers at different locations \cite{Storey94}. The complete calculation of $\Delta\phi_\textrm{propagation}$ is done in Section \ref{sensitivity to B} and shows that $\Delta\phi_\textrm{propagation} = \vec{k}_\textrm{eff} \cdot \vec{a}_B T^2~ +$ smaller terms (see Eq.(\ref{phase 1 et 2})) with $\vec{a}_B$ an acceleration due to a magnetic force. This magnetic acceleration depends on the transition $\pm k_\textrm{eff}$ and is expressed as $\vec{a}_{B}=-\frac{\mu_B}{m} g_F m_F  \vec{\nabla} B =  -\frac{\hbar}{m} \alpha_m \vec{\nabla} B $, where $m$ is the $^{87}$Rb atomic mass, $\hbar$ is the reduced Planck constant, $\alpha_m = \pm 2 \pi \times 0.70$ MHz/G is the Zeeman shift of $\ket{F=1,m_F=\mp1}$ respectively \cite{Steck}, and $\vec{\nabla} B$ is the gradient of the magnetic field magnitude $B$. In the infinitely short, resonant-pulse limit, the second phase term of $\Delta \phi$ is given by $\Delta \phi_\textrm{laser} =\vec{k}_\textrm{eff} \cdot \vec{a}_\text{inertial}~ T^2$ where $\vec{a}_\text{inertial}$ is the  acceleration of the atoms due to gravito-inertial effects. This interferometer geometry thus exhibits at its output an atomic phase shift sensitive to the combined acceleration of the atoms due to gravito-inertial effects and a force due to a magnetic field gradient:

\begin{equation} \label{eq2}	
\begin{aligned}
\Delta \phi &= \vec{k}_\textrm{eff} \cdot \vec{a} ~ T^2 \\
  &=  \vec{k}_\textrm{eff} \cdot \left( \vec{a}_\textrm{inertial} - \frac{\hbar}{m} \alpha_m \vec{\nabla} B \right) T^2
\end{aligned}
\end{equation}

At the end of the interferometric sequence we measure the proportion of atoms in each output port of the interferometer $\ket{F=1}$ and $\ket{F=2}$ by fluorescence. The normalized population in the state $\ket{F=2}$ at the LPAI exit is a sinusoidal function of the interferometer phase shift:

\begin{equation} \label{eq3}
P = P_m - \frac{C}{2} \cos \Delta \phi = P_m - \frac{C}{2} \cos (\vec{k}_\textrm{eff} \cdot \vec{a}~ T^2)
\end{equation}

\noindent where $P_m$ is the fringe offset and $C$ is the fringe contrast. In the following we neglect the smaller terms from $\Delta\phi_\textrm{propagation}$ (see Eq.(\ref{phase 1 et 2})) which will be studied in Section \ref{sensitivity to B}.

The force responsible of the magnetic acceleration depends on the magnetic field inhomogeneities. To evaluate this force in our setup, we proceed as follows: just as in a LPAI gravimeter we apply a radio frequency chirp $\beta$ (Hz/s) to the effective Raman frequency to scan the interference fringes. The sinusoidal dependence of the probability (see Eq.(\ref{eq3})) leads to an ambiguity in the acceleration measurement. We solve this issue by measuring interference fringes for different interrogation times $T$. Reversing the sign of the effective wave vector (\textit{i.e.} preparing the atoms in $\ket{F=1,m_F=\mp1}$ alternatively), the magnetic acceleration changes sign and the phase shifts are respectively:

\begin{equation}
\begin{aligned}
\Delta \phi_\pm &=\left[ \pm \vec{k}_\textrm{eff} \cdot \left( \vec{a}_\textrm{inertial} + \vec{a}_{B\pm} \right) - 2 \pi \beta \right] T^2 \\
&=\left[ \pm \vec{k}_\textrm{eff} \cdot  \vec{a}_\textrm{inertial} + \vec{k}_\textrm{eff} \cdot \vec{a}_{B+}  - 2 \pi \beta \right] T^2
\end{aligned}
\end{equation}

\noindent where 

\begin{equation*} 
	\left\lbrace
\begin{aligned}
 \vec{a}_{B\pm} &= \vec{a}_B \left(F=1,m_F=\mp 1 \right) = \vec{a}_B \left(F=2,m_F=\pm 1\right) \\
\vec{a}_{B+} &= - \vec{a}_{B-}
\end{aligned}
\right.
\end{equation*}

This means that whatever the pulse separation $T$, the phase shift is zero when the chirp reaches $\beta_{0\pm} = \frac{k_\textrm{eff}}{2 \pi} \cdot (\pm a_\textrm{inertial} + a_{B+})$. From this we easily extract the magnetic acceleration $a_{B+} = \frac{2 \pi}{2 k_\textrm{eff}}\cdot (\beta_{0+} + \beta_{0-})$. Its numerical value is $-7.79 \times 10^{-3}$ m.s$^{-2}$, \textit{i.e.} several interfringes $i = \lambda/2T^2 = 1.43 \times 10^{-3}$ m.s$^{-2}$ ($T=16.5$ ms). The corresponding magnetic field gradient is $\partial_x B = - 24.2$ mG.cm$^{-1}$ and is therefore responsible for a non negligible bias on the inertial acceleration measurement. The $k_\textrm{eff}$-reversal technique (\textit{i.e.} preparing the atoms in the $\ket{F=1,m_F=\mp1}$ states alternatively) is essential to eliminate such a bias.

\subsection{Correlation fringes} \label{fringes}

\begin{figure}[h]
\centering
\includegraphics[scale=0.3]{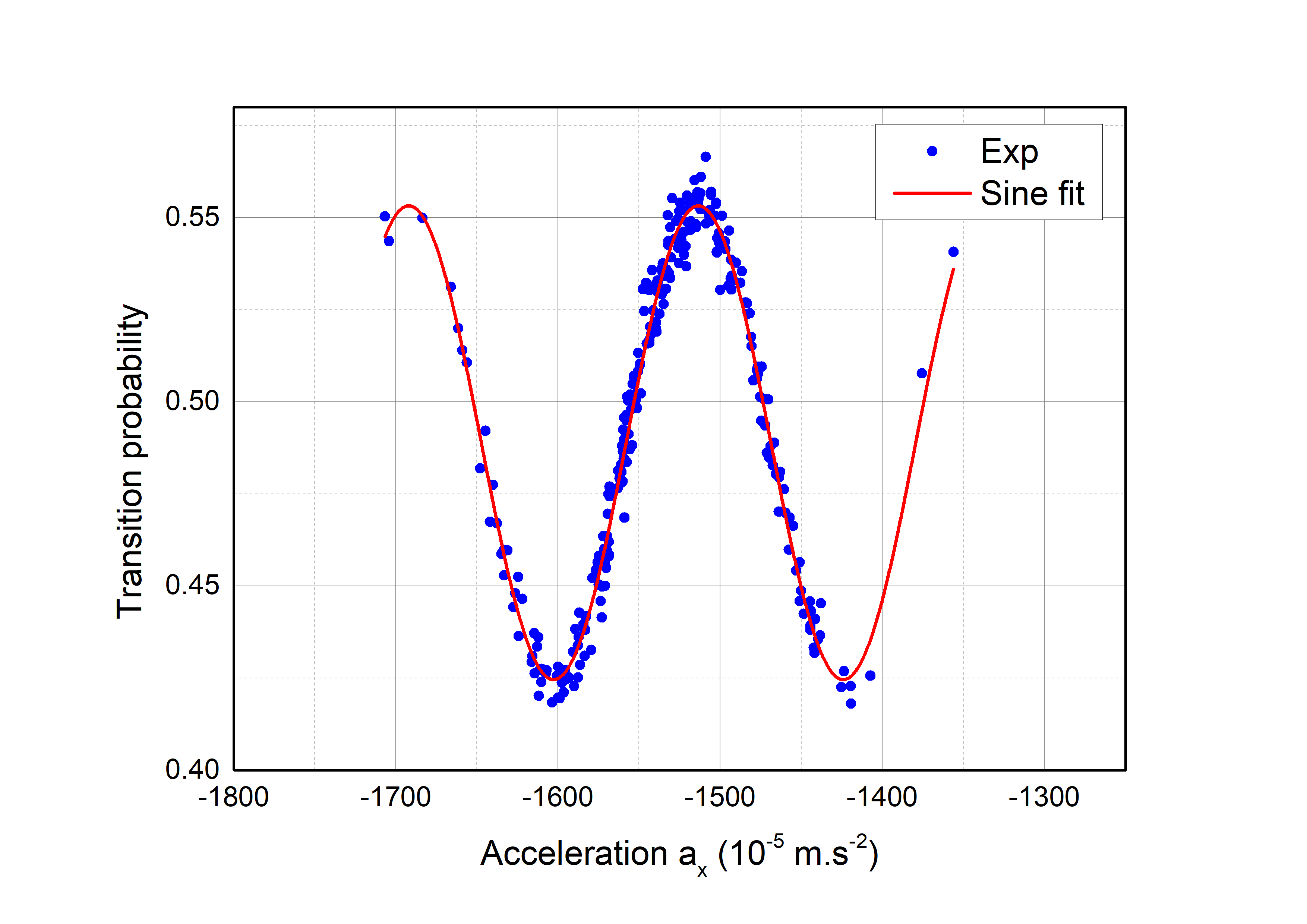}
\caption{Horizontal atom interferometer fringe pattern. The total interferometer time is $2T = 32$ ms and the Raman pulse duration is $\tau=10~\mu$s. The solid line is a sinusoidal least-squares fit using Eq.(\ref{eq3}). The estimated fringe contrast is $C \sim 13~\%$. The fringes are shifted from zero because of the magnetic acceleration and the bias of the classical accelerometer. }
\label{figure4}
\end{figure}

When scanning the interference fringes by varying the frequency chirp $\beta$, the fringes are washed out because of vibration noise (typically as soon as $T \geq 6$ ms). To recover the interference fringes, we perform a correlation-based technique \cite{Lautier2014} which combines both measurements of the LPAI output signal $P$ and of the classical accelerometer fixed to the Raman mirror. Figure \ref{figure4} shows the typical fringe pattern obtained by plotting the transition probability $P$ at the LPAI output versus the acceleration measured by the classical accelerometer. The fringe contrast obtained from a sinusoidal least-squares fit of the data is $C = 13~\%$, which is the best contrast that we obtained when adjusting the Raman laser intensity at a fixed Raman pulse duration of $\tau=10~\mu$s. We demonstrated in \cite{Bernard2019} a horizontal hybrid accelerometer with a contrast of $40~\%$ on the same experimental setup. In Section \ref{specifics} we investigate the loss of contrast associated to the $\sigma^+ - \sigma^-$ technique.

\subsection{Accelerometer sensitivity} \label{sensitivity}

\begin{figure}[h]
\centering
\includegraphics[scale=0.3]{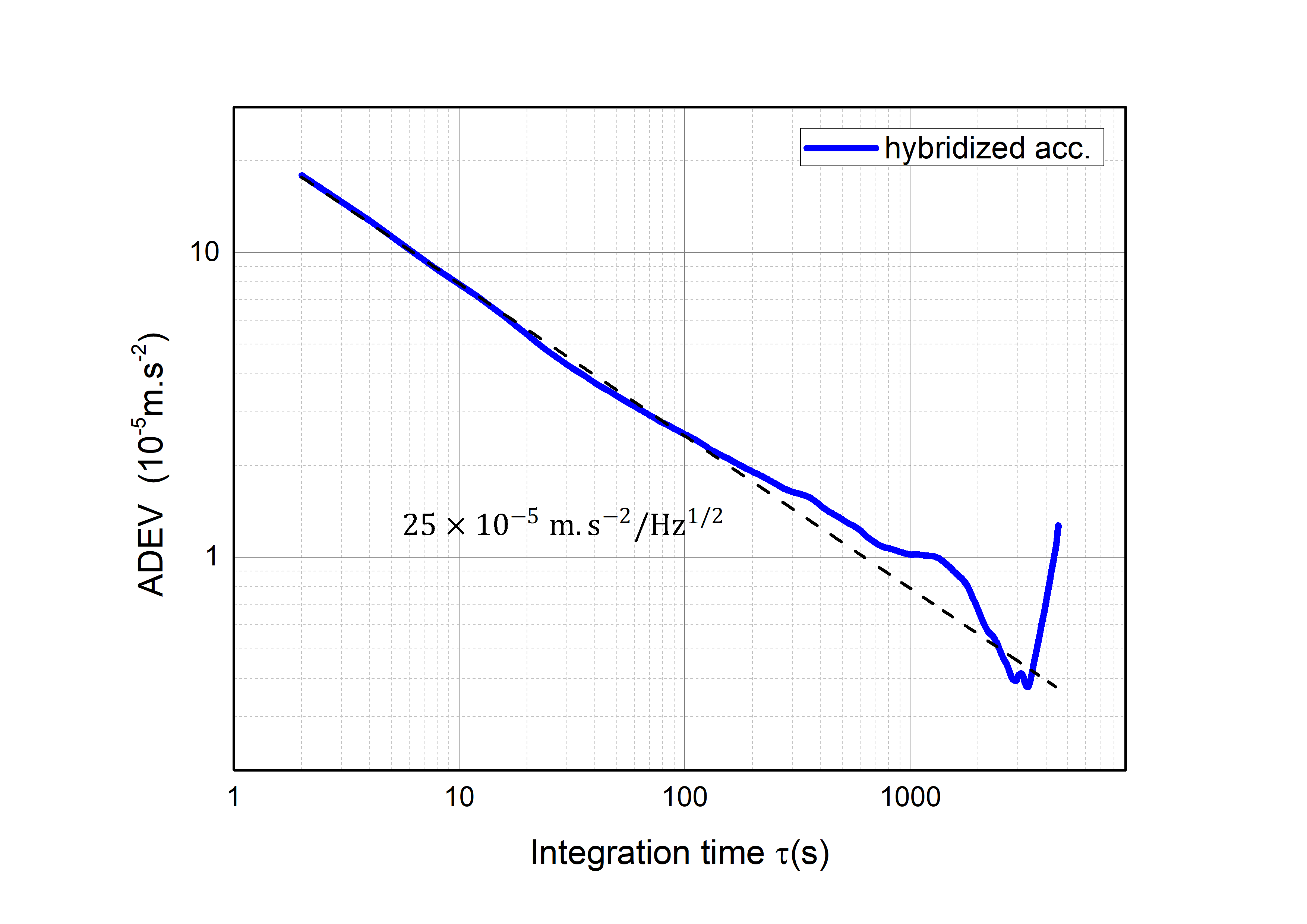}
\caption{Allan standard deviation (ADEV) of the hybridized horizontal atom accelerometer (blue line). The dashed line illustrates the $\tau^{-1/2}$ scaling.}
\label{figure5}
\end{figure}

We analyze the sensitivity and stability of the horizontal atom accelerometer by hybridizing the classical and the atomic sensors \cite{Lautier2014,Bidel2018}. The sign of the effective Raman wave vector $\vec{k}_\textrm{eff}$ is reversed every measurement cycle, \textit{i.e.} the atoms are alternatively prepared in the $\ket{F=1,m_F=\mp1}$ states. We calculate the inertial acceleration by computing the half sum of the phase shifts measured on each correlation fringe pattern $\pm k_\textrm{eff}$. The fringe ambiguity is removed by assuming that the magnetic acceleration has the same value as calculated for smaller interrogation times, \textit{i.e.} $a_B=-7.79 \times 10^{-3}$ m.s$^{-2}$. Figure \ref{figure5} displays the Allan standard deviation (ADEV) of the hybridized atom interferometer signal. We achieve a short-term sensitivity of $25\times 10^{-5}$ m.s$^{−2}$/$\sqrt{\textrm{Hz}}$, which is not as good as the state of the art for horizontal configurations \cite{Xu2017,Bernard2019}. In Section \ref{specifics} we discuss several arguments to explain this sensitivity. The ADEV of the atomic sensor scales as $\tau^{-1/2}$ and reaches $3.8 \times 10^{-6}$ m.s$^{−2}$ after 3300 s integration time. No conclusion can be drawn regarding the long-term stability of the atom accelerometer because of angular drifts of the Raman mirror. An auxiliary tilt sensor could be used to monitor the angle between the Raman beam and the horizontal direction during the measurements.

\section{Specifics of the method} \label{specifics}

\subsection{Spontaneous emission}\label{ES}

\begin{figure}[h]
\includegraphics[trim =1.5cm 0cm 0cm 0cm, clip, scale=0.35]{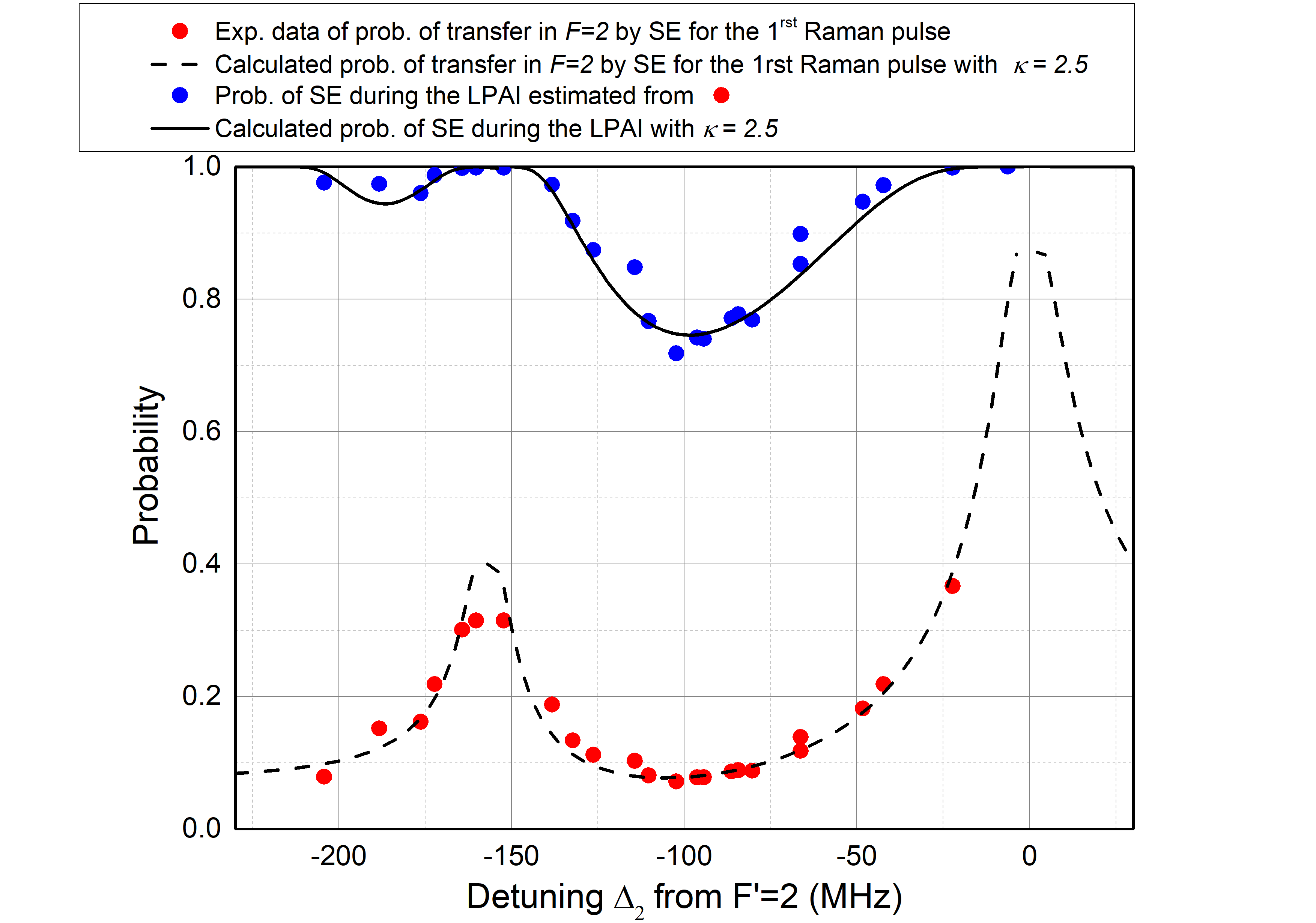}
\caption{Probability of spontaneous emission (SE) as a function of the detuning $\Delta_2$ from the excited state $\ket{F'=2}$. Red dots represent the experimental data of probability of transfer by spontaneous emission in the $\ket{F=2}$ level during the first Raman pulse of duration 10 $\mu$s. The data was obtained from the detected background on Raman spectra plotted for different detunings $\Delta_2$ at constant effective Rabi frequency. The dashed line represents the theoretical probability of transfer by spontaneous emission (adjusted with a parameter $\kappa=2.5$) in the $\ket{F=2}$ level during a Raman pulse of 10 $\mu$s. Blue dots represent the total probability of spontaneous emission in the $\ket{F=1}$ and $\ket{F=2}$ levels during the whole interferometer. It was derived from the experimental data in red dots. The black line is the theoretical total probability of spontaneous emission during the whole interferometer (corrected by a factor $\kappa=2.5$).}
\label{figure6}
\end{figure}

As shown in Eq.(\ref{eq1}) the transition probability amplitudes for each excited state $\ket{F'=1}$ and $\ket{F'=2}$ interfere destructively. In order to address the counterpropagating transition, the laser detuning is set between the $\ket{F'=1}$ and $\ket{F'=2}$ levels and therefore induces spontaneous emission and coherence loss. Assuming that atoms which undergo spontaneous emission do not interfer anymore, the contrast at the atom interferometer output is reduced, and so is the sensor sensitivity. We experimentally tuned the Raman laser frequency to minimize the spontaneous emission rate. The detuning $\Delta_2$ from the excited state $\ket{F'=2}$ is adjusted via a beat-note between the fiber laser and the Raman laser diode. The probability of transfer by spontaneous emission in $\ket{F=2}$ is estimated by measuring the transfer probability during a 10 $\mu$s out-of-resonance Raman pulse. Figure \ref{figure6} shows the experimental result (red dots) as a function of the detuning $\Delta_2$, along with the theoretical prediction of the probability of transfer by spontaneous emission in the $\ket{F=2}$ level during a 10 $\mu$s pulse. In order to match the experimental result with the theoretical prediction, we introduce an empirical multiplicative factor $\kappa=2.5$ for the laser intensity (see Supplemental Material at [...] for detailed calculations).
Considering the dipole matrix elements we can derive an estimation of the total spontaneous emission probability during the whole interferometer both from data and theory (blue dots and black line in Fig.\ref{figure6}). Here again we assume $\kappa=2.5$ in the theoretical calculations. The need of this empirical factor $\kappa$ is not understood. It could come from experimental defaults (non perfect $\pi/2$ Raman pulses) and from our theoretical treatment of spontaneous emission that is too simple (the resolution of Bloch optics equations could be used in a more elaborated model \cite{Cheng2016}).

The optimal detuning is given by the curve minimum: $\Delta_2 = -100$ MHz both theoretically and experimentally. We can conclude from Figure \ref{figure6} that for $\Delta_2=-100$ MHz, 70 $\%$ of the atoms undergo spontaneous emission during the interferometer duration. Such a loss of atoms reduces the contrast by a factor 3 and could explain our relatively low contrast value of $13 \%$ (see Figure \ref{figure4}). As a matter of fact, we demonstrated in \cite{Bernard2019} a horizontal hybrid accelerometer on the same experimental setup with a contrast of $40 \%$. Since $ 40 \% \times 1/3 \simeq 13 \%$, the visiblity loss is probably due to spontaneous emission. Nevertheless, we investigate in the next subsection the phase shift sensitivity to magnetic field and show how it also affects the LPAI contrast and bias.

Setting the detuning close to resonance still has an advantage: the transition requires low Raman intensity ($\sim I_\textrm{sat}$) compared to the magnetically insensitive Raman transition ($\sim 10~ I_\textrm{sat}$) for which the detuning is set far from resonance. However spontaneous emission can be drastically reduced if the Raman transition is performed on the $D_1$ line instead of the $D_2$ line of $^{87}$Rb, since the hyperfine levels are further apart. Theoretical calculations of spontaneous emission on the $D_1$ line show that only 10 $\%$ of the atoms undergo spontaneous emission during the interferometer sequence, which is comparable to LPAIs using the $\ket{F=1, m_F=0} - \ket{F=2,  m_F=0}$ transition with a commonly used one-photon detuning from $\ket{F'=3}$ of $\sim -1$ GHz.

\subsection{Sensitivity to magnetic field} \label{sensitivity to B}

As shown in Eq.(\ref{eq2}) the LPAI is sensitive to both inertial acceleration and magnetic forces from field inhomogeneities. Using the $k_\textrm{eff}$-reversal technique one can extract each contribution by computing either the half sum or the half difference of the phase shifts (see Section \ref{AI technique}). This is only valid under the assumption of a constant magnetic field gradient from shot-to-shot. We perform in this Section the detailed calculation of the phase shift due to magnetic field by taking into account spatial inhomogeneities of the magnetic field up to order two. From this study we estimate the bias and the loss of contrast induced by the magnetic field. \\

We have considered so far a weak magnetic field and a linear relationship between magnetic energy levels and magnetic field, with the same shift for $\ket{F=1,m_F=\mp1}$ and $\ket{F=2,m_F=\pm1}$. For the ground state manifold of the $D$ transition, the exact calculation of the energy levels is given by the Breit-Rabi formula \cite{BreitRabi1931}. In the case of $^{87}$Rb in the $\ket{F=1,m_F=- 1}\equiv\ket{1,- 1}$ and $\ket{F=2,m_F= 1} \equiv \ket{2, 1}$ levels, the energies are given by:

%
%
%


\begin{equation} \label{br omega}
	\left\lbrace
\begin{aligned}
 E_{\ket{1,-1}} &= \hbar  \left( \alpha_m - \Delta \alpha \right) \times B \equiv \hbar \omega_a \\
E_{\ket{2,+1}} &= \hbar \left( \alpha_m + \Delta \alpha \right) \times B \equiv \hbar \omega_b
\end{aligned}
\right.
\end{equation}

\noindent $\mathrm{with~~} \alpha_m = \dfrac{g_J - g_I}{4 \hbar }\mu_B  \mathrm{~~and~~}  \Delta \alpha = \dfrac{g_I}{\hbar} \mu_B. $ Here we do  not take into account the hyperfine splitting, as it is a constant which cancels out in the phase shift calculation. \\

\noindent $a,b$ stand for the two hyperfine states $\ket{1,-1}$, $\ket{2,+1}$ of the LPAI. $g_I$ is the nuclear g-factor, $g_J$ is the Land\'e factor and $\mu_B$ is the Bohr magneton. Under the assumption of spins following the magnetic field adiabatically during free fall, $B$ represents the magnitude of the magnetic field. We define the average energy shift $\alpha_m =2 \pi \times  0.70$ MHz/G (see Eq.(\ref{eq2})) and the differential energy shift $\Delta \alpha = 2 \pi \times -1.4$ kHz/G coming from the Breit-Rabi formula. Similarly, for the $-k_\textrm{eff}$ transition between $\ket{1,+1}$ and $\ket{2,-1}$, the energy levels are described by Eq.(\ref{br omega}), only  $(\alpha_m, \Delta \alpha)$ are replaced with $(-\alpha_m,- \Delta \alpha)$. \\

We use the Feynman path integral approach to compute the magnetic phase shift between the two arms $(u,d)$ of the interferometer. Using a perturbative calculation for the effect of the magnetic field \cite{Storey94}, the phase $\varphi^{(i)}$ accumulated along each arm $i=u,d$ of the LPAI is given by the classical action $S_\mathrm{cl}$ along the unperturbated classical path divided by $\hbar$:

\begin{equation} 
\varphi^{(i)} = S^{(i)}_\mathrm{cl} /\hbar =  \int_0^{2T} \mathcal{L}[\vec{r}(t)] /\hbar \;\mathrm{d}t
\end{equation} 

\noindent The phase difference at the output of the LPAI is then:

\begin{equation} \label{deph}
\begin{aligned}
\Delta \phi_\textrm{path} &= \varphi^{(u)}-\varphi^{(d)} \\
&= (S^{(u)}_\mathrm{cl}-S^{(d)}_\mathrm{cl})/\hbar \\
 &= 1/\hbar \left( \int_0^{2T} \mathcal{L}[\vec{r}~^{(u)}(t)]\;\mathrm{d}t -\int_0^{2T} \mathcal{L}[\vec{r}~^{(d)}(t)]\;\mathrm{d}t \right) \\
&= \int_0^{T} \left[ \omega_b \left( \vec{r}~^{(u)}(t)\right) - \omega_a \left( \vec{r}~^{(d)}(t) \right) \right]\;\mathrm{d}t \\
&+ \int_T^{2T} \left[ \omega_a \left(\vec{r}~^{(u)}(t) \right) - \omega_b \left( \vec{r}~^{(d)}(t) \right) \right] \;\mathrm{d}t \\
\end{aligned}
\end{equation}

Taking into account the Breit-Rabi correction (see Eq.\ref{br omega}), the phase difference at the output of the interferometer can be split into two terms:

\begin{equation} \label{2 termes}
\begin{aligned}
\Delta \phi_\textrm{path}  &= \alpha_m \int_0^{2T} \left[ B \left( \vec{r}~^{(u)}(t) \right) -B \left( \vec{r}~^{(d)}(t) \right)  \right] \;\mathrm{d}t \\
&+\Delta \alpha \left( \int_0^{T} \left[ B \left( \vec{r}~^{(u)}(t) \right) + B \left( \vec{r}~^{(d)}(t) \right)  \right] \right.\\
 &- \int_T^{2T} \left[ B \left( \vec{r}~^{(u)}(t) \right) + B \left( \vec{r}~^{(d)}(t) \right)  \right] \;\mathrm{d}t   \Bigg. \Bigg) 
 \end{aligned}
\end{equation}

The first term (proportional to $\alpha_m$) arises from the magnetic field variation between the upper and the lower arms of the interferometer, whereas the second term (proportionnal to $\Delta \alpha$) is accounting for the variation of the mean field $B$ between $[0-T]$ and $[T-2T]$. \\

The phase shift calculation is performed on the unperturbed trajectories $ \left( \vec{r}~^{(u)}(t),\vec{r}~^{(d)}(t) \right)$ whose expressions are:

\begin{equation} \label{r}
\left\lbrace
\begin{aligned}
\vec{r}~^{(u)}(t) &= \vec{r}_0 + (\vec{v}_0 + \vec{v}_\textrm{eff}) t - \vec{v}_\textrm{eff} (t-T) \mathcal{H} (t-T) + \frac{1}{2} \vec{g} t^2 \\
\vec{r}~^{(d)}(t) &= \vec{r}_0 + \vec{v}_0 t + \vec{v}_\textrm{eff} (t-T) \mathcal{H} (t-T) + \frac{1}{2}\vec{g} t^2 
\end{aligned}
\right.
\end{equation}

\noindent where $\vec{v}_\textrm{eff} = \hbar \vec{k}_\textrm{eff} /m$, $(\vec{r}_0,\vec{v}_0)$ are the position and velocity vectors at the first Raman pulse and $\mathcal{H}(t)$ is the Heaviside function. \\

The magnetic field magnitude $B$ is supposed to be time-independent and can therefore be expressed through its Taylor expansion in space:

\begin{equation} \label{B}
\begin{aligned}
B(\vec{r}) &= B_0 + \vec{B}_1 \cdot \vec{r} + \frac{1}{2} \vec{r} \cdot \bar{\bar B}_2 \cdot  \vec{r} \\
&= B_0 + \partial_xB x + \partial_y B y + \partial_z B z + \frac{1}{2} ( \partial_x^2B x^2 + \partial_y^2 B y^2 \\
&+ \partial_z^2 B z^2 + 2 \partial_x \partial_y B xy + 2 \partial_x \partial_z B xz + 2 \partial_y \partial_z B yz )
\end{aligned}
\end{equation}

Using Eq.(\ref{r},\ref{B}) the phase shift calculation of Eq.(\ref{2 termes}) leads to two terms proportional to $\alpha_m$ and $\Delta \alpha$ respectively:

\begin{equation}
\Delta \phi_\textrm{path} = \Delta\phi_1 + \Delta\phi_2
\end{equation}

\noindent where

\begin{equation} \label{phase 1 et 2}
\left\lbrace
\begin{aligned}
\Delta \phi_1 = \alpha_m T^2 &\Big( \Big. \partial_xB v_\textrm{eff}   \\
&+  \partial_x^2B v_\textrm{eff}  \left[ x_0 + ( v_{0x} + v_\textrm{eff}/2 ) T\right]\\
& +  \partial_x \partial_yB v_\textrm{eff}  \left[ y_0 + v_{0y} T \right]  \\  &+  \partial_x \partial_z B v_\textrm{eff}\left[z_0 + v_{0z} T + 7/12 g T^2 \right]  \Big. \Big) \\
\Delta \phi_2 = -\Delta \alpha T^2 &\Big( \Big. 2  \partial_x B (v_{0x} + v_\textrm{eff}/2)  \\
&+ 2 \partial_y B v_{0y}  \\ &+ 2  \partial_z B (v_{0z} + gT)  \Big. \Big)
\end{aligned}
\right.
\end{equation} 

The first term of $\Delta \phi_1$ is the magnetic acceleration of Eq.(\ref{eq2}) since $\alpha_m \partial_xB v_\textrm{eff} T^2  = \alpha_m \partial_x B \frac{\hbar}{m} k_\textrm{eff} T^2 = k_\textrm{eff} a_B T^2$. The other terms of $\Delta \phi_1$ are due to magnetic field curvatures whereas the terms of $\Delta \phi_2$ come from the differential shift $\Delta \alpha$ and the magnetic field gradient. From Eq.(\ref{phase 1 et 2}) one can deduce the bias and loss of contrast induced by the $\sigma^+-\sigma^-$ method. \\

The $k_\textrm{eff}$-reversal technique enables to eliminate any systematic effect whose sign does not change when reversing $\vec{k}_\textrm{eff}$. The bias is then due to the remaining phase terms. In our case, since $(\alpha_m, \Delta \alpha)$ change sign when reversing $\vec{k}_\textrm{eff}$, we deduce from Eq.(\ref{phase 1 et 2}) that the bias induced by magnetic effects is:

\begin{equation} \label{bias}
 \frac{1}{2}\alpha_m \partial_x^2 B v_\textrm{eff}^2 T^3 -
 2 \Delta \alpha \partial_z B g T^3
\end{equation}

The first term is an inertial phase due to the atom recoil and the magnetic field curvature: it arises as soon as the magnetic field gradient is different for the upper and lower arm of the interferometer. It was estimated theoretically by computing the magnetic field produced by the horizontal field coils ($\partial_x^2B \sim 5$ G.m$^{-2}$) and leads to a bias of $2\times 10^{-6}$ m.s$^{-2}$ (we recall $T=16.5$ ms). The second term is an energy dependant phase term which comes from time variation of the differential Zeeman energy shift induced by the magnetic field gradient $\partial_zB$ seen by the atoms during free fall. In order to estimate it we measured the magnetic field vertical gradient by Zeeman spectroscopy and found $\partial_z B = 1.2$ G.m$^{-1}$. The resulting bias is $2 \times 10^{-4}$ m.s$^{-2}$. In comparison, ref.\cite{Wodey2020} demonstrates a high-performance magnetic shield for long baseline atom interferometry with inhomogeneities below 3 nT/m: the associated bias would be $3 \times 10^{-9}$ m.s$^{-2}$. From Eq.(\ref{phase 1 et 2}) one can easily notice that this bias can be eliminated through an atomic fountain design with a properly set vertical velocity at the first Raman pulse $v_{0z} = -gT$. One can also set the quantization field to its magic value \cite{Cornell2002} corresponding to the magnetic field at which the derivative of the energy difference $\hbar (\omega_b - \omega_a)$ is null. But this configuration does not cancel out the bias perfectly because it requires to change the magnetic field sign when changing the sign of  $\vec{k}_\textrm{eff}$, since the magic field for the $\pm \vec{k}_\textrm{eff}$ transitions are respectively $\pm 3.2$ G \cite{Cornell2002}.   \\

\begin{figure}[h]
\centering
\includegraphics[scale=0.4]{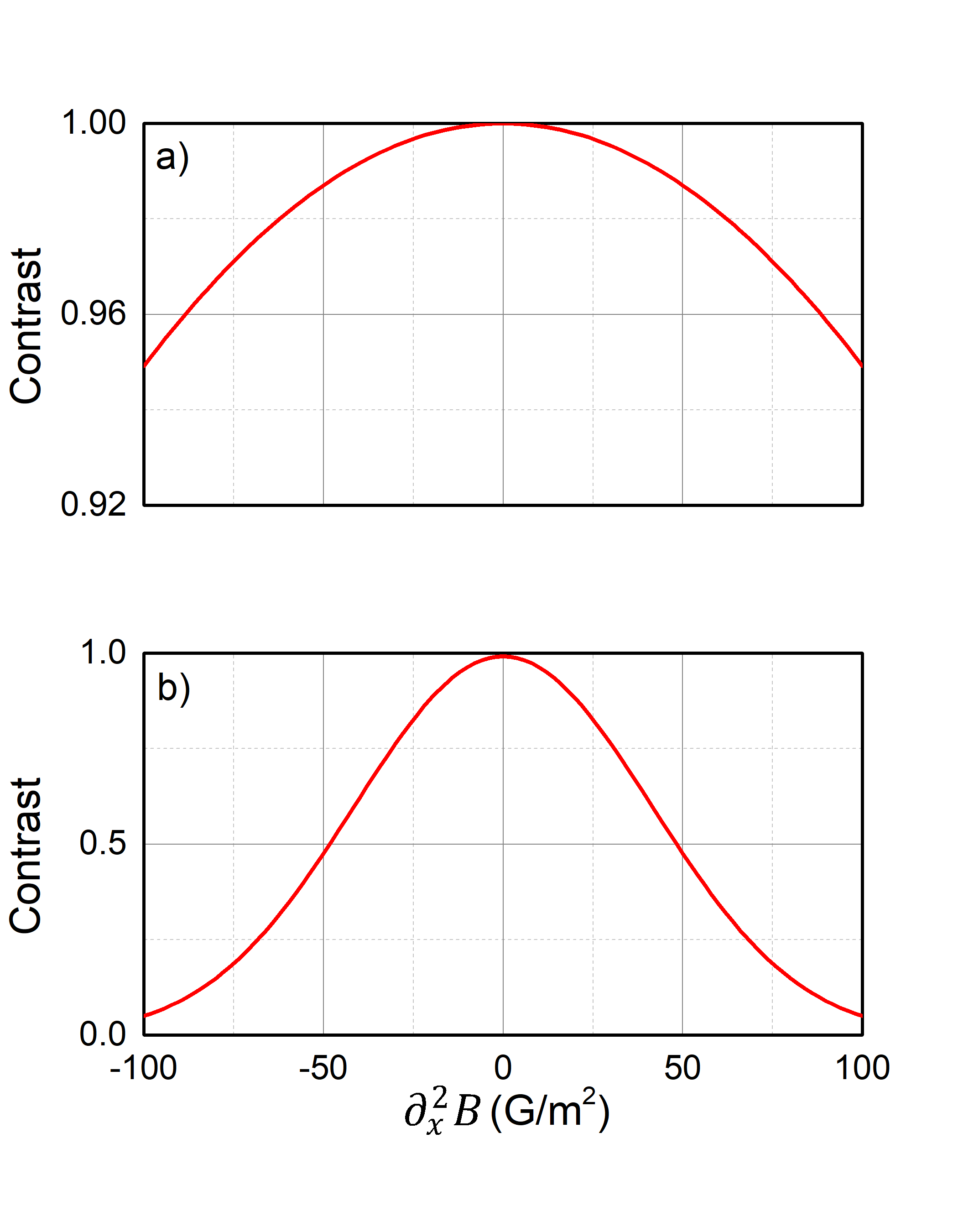}
\caption{Theoretical contrast calculated theoretically as a function of the magnetic field curvature. Parameters: interrogation time $T=16.5$ ms, gas temperature $\theta=2~\mu$K, typical atomic cloud size $\sigma_x = 1$ mm, velocity $\sigma_v = \sqrt{k_B \theta/m}=1.4$ cm.s$^{-1}$. (a) Contrast loss due to the velocity dependent phase shift $\alpha_m \partial_x^2 B v_\textrm{eff} v_{0x} T^3$. (b) Contrast loss due to the position dependent phase shift $\alpha_m \partial_x^2 B v_\textrm{eff} x_0 T^2$. }
\label{figure7}
\end{figure}

Regarding the contrast reduction induced by magnetic effects, it is due to the position and velocity dependent terms in Eq.(\ref{phase 1 et 2}). Indeed any phase shift sensitivity to position (or velocity) results in each atom (or velocity class) providing its own fringe pattern. Since the atom detection protocol averages over these patterns, the fringe contrast is reduced. In our case the interrogation time $T=16.5$ ms is short enough to neglect the loss of contrast arising from velocity dependant phase shifts in Eq.(\ref{phase 1 et 2}) (see Fig.\ref{figure7}(a)). The contrast reduction due to averaging over position on the other hand is much more significant. As an example we analyse  the phase shift $\alpha_m \partial_x^2 B v_\textrm{eff} x_0 T^2$: it reduces contrast by a factor $e^{-\frac{1}{2}(\alpha_m \partial_x^2 B v_\textrm{eff} T^2 \sigma_x)^2}$ \cite{Roura2014}, where $\sigma_x = 1$ mm is the typical size of the atomic cloud. Fig.\ref{figure7}(b) illustrates this loss of contrast as a function of the magnetic field curvature $\partial^2_x B$: the contrast is typically reduced by 50 $\%$ in the presence of $\partial^2_x B \sim 50$ G.m$^{-2}$. As comparison Fig.\ref{figure7}(a) shows how the velocity dependent phase shift $\alpha_m \partial_x^2 B v_\textrm{eff} v_{0x} T^3$ does not affect contrast. Since the magnetic field along the Raman beams cannot be measured precisely enough on our experimental setup, we estimated theoretically the curvature due to our magnetic field coils and found  $\partial_x^2B_\textrm{th} \sim 5$ G.m$^{-2}$. From this we can state that inhomogeneities of the magnetic field created by the coils do not affect the fringe contrast. However the presence of another magnetic field source creating a non negligible field curvature responsible of a contrast reduction is to be considered.  \\

Loss of contrast can also be interpreted as in \cite{Roura2014}: position and velocity dependent phase terms in Eq.(\ref{phase 1 et 2}) are responsible for the opening of the interferometer in momentum and position respectively. We introduce the notation $\Gamma = 2 \frac{\hbar}{m} \alpha_m \partial_x^2 B$ since the magnetic field curvature is the exact analog of a gravity gradient in a vertical LPAI. The forces associated with $\Gamma$ tend to open up the trajectories of the atoms and lead to an open interferometer with nonvanishing relative position and momentum displacements at the output of the LPAI. As demonstrated in \cite{Roura2014} a position dependent phase shift results in a momentum displacement $\delta P$ at the output of the LPAI, and a velocity dependent phase shift results in a position displacement $\delta X$. Both displacements are given by the following expressions to first order in $\Gamma T^2$ \cite{Roura2014}: 

\begin{align}
\delta X &= \left( \Gamma T^2 \right) \frac{\hbar k_\textrm{eff}}{m} T  \label{deltax}\\
\delta P &= \left( \Gamma T^2 \right) \hbar k_\textrm{eff} \label{deltap}
\end{align}

In our case the change of position associated with Eq.(\ref{deltax}) is very small compared to the coherence length which is estimated by the thermal De Broglie wavelength: $\delta X \sim 10^{-8}$ m $\ll l_c= \lambda_\textrm{DB} = h/\sqrt{2 \pi m k_B \theta} \sim 10^{-7}$ m for $^{87}$Rb atoms at $\theta=2~\mu$K. On the other hand the momentum displacement given by Eq.(\ref{deltap}) is not as negligible as the position displacement. Even though \cite{Roura2017} suggests a protocol to mitigate the loss of contrast due to gravity gradients $\Gamma$ through a suitable adjustment of the laser wavelength at the second Raman pulse, this technique cannot be applied here. Indeed the $\sigma^+ - \sigma^-$ method requires the detuning $\Delta$ to stay between the $\ket{F'=1}$ and $\ket{F'=2}$ sublevels to avoid spontaneous emission (see Section \ref{ES}), which means that the mitigation technique proposed by \cite{Roura2014} would inevitably reduce the fringe contrast.

\subsection{Light shifts}

In most cases one can eliminate the differential one-photon light shift by adjusting the intensity ratio between the two Raman lasers. In our $\sigma^+ - \sigma^-$ Raman transition scheme, there is no intensity ratio that cancels out the differential one-photon light shift (see Supplemental Material at [...] for complete calculation of light shifts), which means that an intensity variation between the first and the last pulses of the LPAI leads to a residual parasitic phase shift. For an intensity variation of 10 $\%$ between the first and the last pulses, the corresponding bias due to the one-photon light shift is $-1.5 \times 10^{-5}$ m.s$^{-2}$. However this light shift is nearly rejected through the $k_\textrm{eff}$-reversal technique, since $\delta_{\textrm{LS1}}^\textrm{diff}(+k_\textrm{eff}) \simeq \delta_{\textrm{LS1}}^\textrm{diff} (-k_\textrm{eff})$. More generally it is essential to note that the one-photon light shift doesn't represent a limit to our $\sigma^+ - \sigma^-$ technique, since one can find an intensity ratio canceling it out on the $D_1$ line of $^{87}$Rb.

In our setup, the two-photon light shift arises from the off-resonant copropagating Raman transitions detuned by $ \pm \dfrac{2 \mu_B g_F}{h}B$ from the considered $\pm k_\textrm{eff}$ counterpropagating transition. Its effect decreases with the magnetic field. When $B=400$ mG, it is calculated to be negligible compared to the one-photon light shift and it cancels out as well when the $k_\textrm{eff}$-reversal technique is applied. \\

\section{Applications}
In this section we propose some possible applications of our technique in both metrology and for development of new designs of atom interferometers dedicated to field applications. 

\subsection{ $h/m_X$ measurement}
Atom interferometers allow to determine the fine-structure constant $\alpha$ based on measuring the recoil velocity $v_r=\hbar k/m_X$ of an atom $X$ of mass $m_X$ absorbing a photon of momentum $\hbar k$, where $\hbar=h/2\pi$, and $k$ is the photon wave number. With an accurate measurement of $k$, $h/m_X$ can me measured and $\alpha$ can be determined allowing to test the standard model and beyond \cite{Muller2018,Morel2020}.

Here, using our  interferometric design we propose to measure  $h/m_X$ by  combining the interferometric measurement of the magnetic acceleration $a_{B}$  and an independent measurement of the magnetic field gradient $\partial_xB$ thanks to a micro-wave spectroscopy. 
In Eq.(\ref{eq2}), using the $k_\textrm{eff}$-reversal technique, one can isolate the magnetic acceleration, leading to:
\begin{equation}
\frac{h}{m_X}=2 \pi \frac{a_{B}}{\alpha_m \partial_xB}
\end{equation}
Considering state-of-the-art atom accelerometers at their best level of accuracy  ($\sim 10^{-8}$ m.s$^{-2}$) \cite{Karcher2018}, combined with a micro-wave Ramsey interferometer with a free-evolution time $T$=10 ms and a signal-to-noise ratio SNR $=10^3$, one could measure magnetic fields at the level of $\sim 100$ nG. Thus, under an acceleration $a_{\mathrm{B}}=1$ m.s$^{-2}$ one could obtain a relative uncertainty at the level of $10^{-8}$ on $h/m_X$ measurement.

\subsection{Force-balanced atom accelerometer}

\begin{figure}[h]
\centering
\includegraphics[scale=0.6,trim= 4cm 2cm 0cm 4cm,clip]{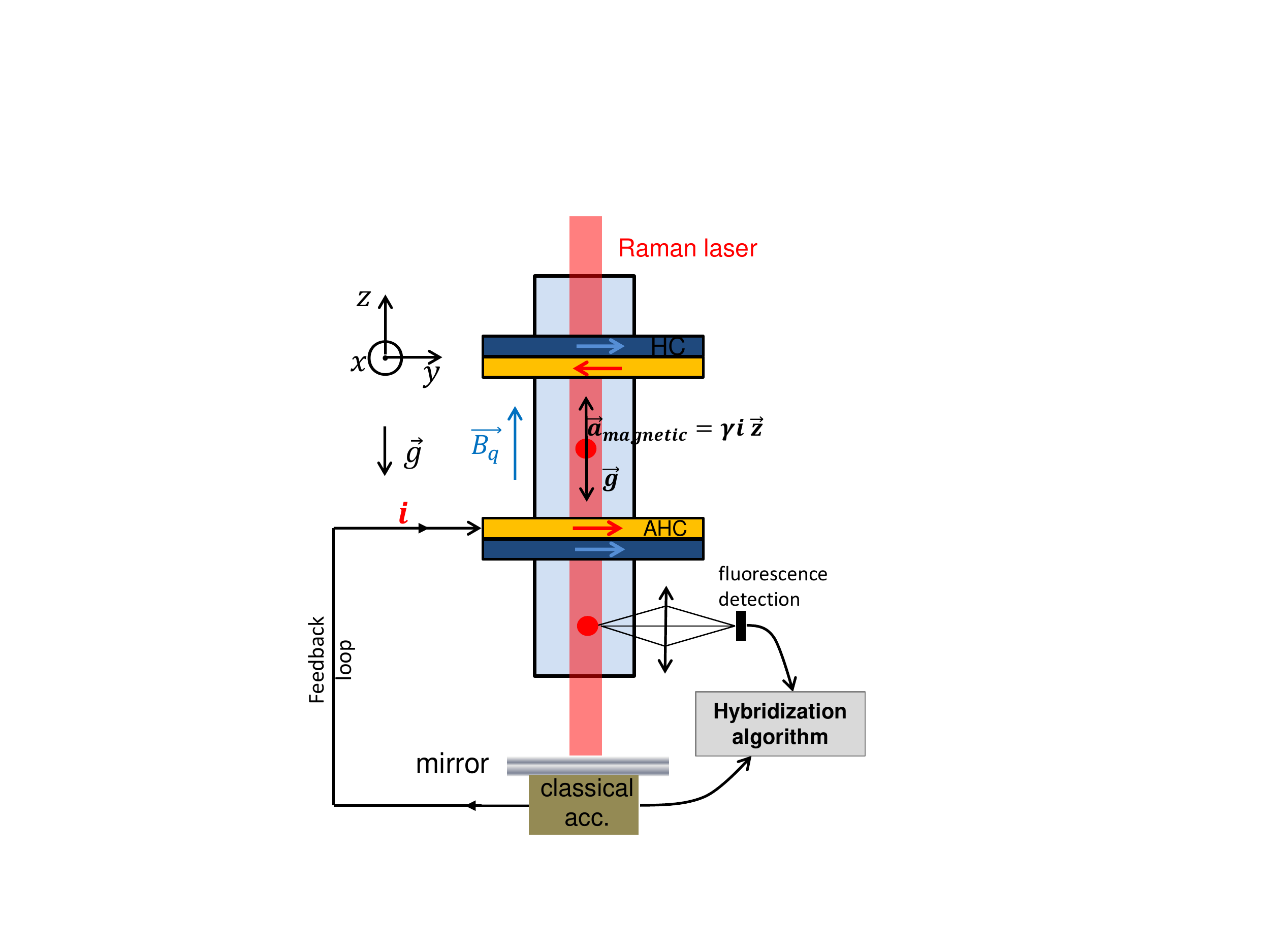}
\caption{Scheme of the force-balanced accelerometer.  AHC: Anti-Helmholtz Coil ; HC: Helmholtz Coil.}
\label{figure8}
\end{figure}

The supplementary internal degree of freedom provided by the magnetically-sensitive states used in the interferometer allows  to exploit the sensitivity of the atoms to magnetic field gradients and transfer a magnetic acceleration onto them. The magnetic acceleration of a Rb atom in a $m_F=\pm 1$ sublevel due to a magnetic field gradient $\partial_z B$ (T.m$^{-1}$) is $a_{B}\sim 32.1\times \partial_z B$ m.s$^{-2}$.
For example, applying a magnetic field gradient of magnitude $\partial_z B\sim 30$ G.cm$^{-1}$ could compensate for gravity acceleration on Earth. Thus, we propose to use our interferometric scheme to create a force-balanced atom accelerometer where the inertial acceleration undergone by the atoms, and measured by an auxiliary classical accelerometer, could be compensated by applying a magnetic acceleration.
The basic principle of the technique is depicted in Fig.(\ref{figure8}).

The atom accelerometer is hybridized with a conventional classical accelerometer in order to track the bias of the classical accelerometer such as in \cite{Bidel2020}. Our coil configuration consists of two vertical coils above and below the atoms.
A pair of Helmholtz coils generates a bias field $\vec{B}_q$ along the $z$-axis and defines the quantization axis. A second pair of coils with counterpropagating currents (anti-Helmholtz configuration) in the same housing, is set to create the magnetic field gradient along the same direction. The current output of the classical accelerometer, proportional to the inertial acceleration, is then used as an input signal to counter-balance the inertial force undergone by the atoms. In pratice, depending on the electric current fed into the anti-Helmholtz coils, one can adjust the magnetic acceleration to $a_{B}=\gamma i$ where $i$ is the electric current and $\gamma$ a scale factor which can be precisely determined through calibration of the magnetic acceleration in a lab-based environment. Thus, it is for example possible to levitate the atoms against gravitational acceleration and therefore extend the evolution time in earthbound laboratories.
Additionnally, using three pairs of Helmholtz coils in  3-orthogonal directions, this scheme could be further extended to compensate for any inertial acceleration in the 3 dimensions, where 3 classical accelerometers are fixed to 3 atom accelerometers to form a 3-axis hybrid accelerometer. With this scheme, one could simultaneously apply a magnetic force on the atoms in the 3 dimensions.
This could benefit onboard atom accelerometers submitted to spurious accelerations which limit the dynamic range because the atoms drop out from the laser beams and the detection zone.


%

\section{Conclusion}
We reported on the experimental demonstration of a horizontal cold-atom interferometer  using counterpropagating Raman transitions between $\ket{F=1,m_F=\mp 1}$ and $\ket{F=2,m_F= \pm1}$ of $^{87}$Rb. Using the same $\sigma^+-\sigma^-$  polarized light arrangement as the MOT, we generated Raman coupling between the two states of the interferometer and showed that this technique allows to perform single-diffraction Raman process in a close-to-zero velocity regime without the need for alternative techniques \cite{Bernard2019}.
We demonstrated that this technique presents both advantages and disadvantages compared to the $\ket{F=1, m_F=0} - \ket{F=2,  m_F=0}$ transition usually used in atom interferometry (see TABLE \ref{tab:tableau}).

Absolute horizontal acceleration measurement with a short-term sensitivity of $25 \times 10^{-5}$ m.s$^{-2}$.Hz$^{-1/2}$ was achieved.
In our setup, limitations of the sensitivity arise from  spontaneous emission, leading to a reduction of the interferometer contrast. The accuracy of the atom accelerometer is mainly limited by the bias caused by the magnetic field gradient at the level of $\sim 2\times 10^{-4}$ m.s$^{-2}$.

Although the short-term sensitivity is bigger by almost one order of magnitude in comparison with state-of-the-art horizontal atom accelerometers \cite{Xu2017,Bernard2019}, it could be improved   by changing the Raman excitation scheme to the D$_1$ line, thus reducing spontaneous emission. 
Additionally, one could reduce this acceleration bias at the level of $\sim 3\times 10^{-9}$ m.s$^{-2}$ (with $T=16.5$ ms) considering a high-performance magnetic shielding leading to a magnetic-field inhomogeneity of 3 nT.m$^{-1}$ \cite{Wodey2020}. 
Moreover, we showed that acceleration bias could be suppressed by performing the atom interferometer using a fountain geometry.
Finally, we believe that using the supplementary internal degree of freedom provided by atoms manipulated in magnetically sensitive levels provides interesting features such as levitation schemes for inertial applications requiring compact setups.

\begin{center}
\begin{table}[h]
\begin{tabular}[c]{|p{4 cm}||p{4 cm}|}
\hline
$$\bf{Advantages}$$&$$ \bf{Disadvantages}$$\\
\hline
Same polarization arrangement MOT/Raman & Spontaneous emission \newline (reduced on the D$_1$ line of $^{87}$Rb)  \\
\hline
Single counterpropagating transition in retroreflective geometry  despite zero Doppler shift& Remaining copropagating Raman transition\\
\hline
Supplementary internal degree of freedom provided by adressing $m_F\neq 0$ Zeeman sublevels\newline (force-balanced accelerometer) & Require precise control and mapping of the magnetic field \\
\hline
\end{tabular}
\caption{Summarized advantages and disadvantages of a Mach-Zehnder cold-atom interferometer using counterpropagative Raman transitions between the state $\ket{F=1,m_F=\mp1}$ and $\ket{F=2,m_F=\pm1}$ of $^{87}$Rb. }
\label{tab:tableau}
\end{table}
\end{center}

\begin{acknowledgments}
M. C acknowledges funding from ONERA through research project CHAMOIS (Centrale Hybride AtoMique de l'Onera et Inertie Strapdown).

\end{acknowledgments}

\newpage

\begin{center}
\textbf{\Large{ Supplemental Material}}
\end{center}

\vspace{0.5cm}

In this Supplemental Material we present the theoretical calculation of the probability of transfer by spontaneous emission  and the light shifts featured in the $\sigma^+-\sigma^-$ Raman transition technique.

\setcounter{section}{0}
\section{Probability of transfer by spontaneous emission}

We start by calculating the effective two-photon Rabi frequency $\Omega_\textrm{eff}$ describing the Raman coupling between the hyperfine states $\ket{F=1,m_F=-1}$ and $\ket{F=2,m_F=1}$ of the interferometer. We compute the single-photon scattering rate $R_\textrm{sc}^F$ from atoms starting in $\ket{F=1}$ or $\ket{F=2}$. From these the probability of transfer by spontaneous emission for an entire interferometer can be calculated. \\

We describe the intensity $I_n$ of each EOM sideband $n$ at the output of the phase modulator as $I_n = I \cdot \mathcal{J}_n(\beta)^2$ where $I$ is the total laser intensity, $\mathcal{J}_n$ is the Bessel function of the first kind of order $n$, and $\beta$ is the modulation index of the EOM. Here we only take into account two sidebands ($n=0$ and $n=1$), the others being detuned enough to be neglected. $I_\textrm{sat}$ stands for the saturation intensity and $\Gamma$ is the natural line width. We calculate the effective two-photon Rabi frequency: 

\begin{equation} \label{omegaeff}
\Omega_\textrm{eff} = \dfrac{\Gamma^2}{2}  \dfrac{I}{I_\textrm{sat}} \mathcal{J}_0(\beta) \mathcal{J}_1(\beta) \sum_{F'}  \dfrac{M_{1,-1}^{F',+} \cdot M_{2,1}^{F',-}}{\Delta_{F'}}
\end{equation}

\noindent where $M_{F,m_F}^{F',\pm} = \dfrac{ \bra{F,m_F} er \ket{F',m_F \pm 1}}{\bra{J=1/2 ||er| }\ket{J'=3/2}} $ are the rubidium $D_2$ dipole matrix elements for $\sigma^{\pm}$ transitions, expressed as multiples of $\bra{J=1/2 ||er| }\ket{J'=3/2}$, as given in \cite{Steck}. $\Delta_{F'}= \omega_L - \omega_2^{F'}$ is the detuning of the carrier (of frequency $\omega_L$) relative to the transition $F=2 \rightarrow F'$ (of energy $\hbar \omega_2^{F'}$). \\

Assuming a $\pi/2$ Raman pulse of duration $\tau$ with Rabi frequency $\Omega_\textrm{eff} \cdot \tau = \pi/2$, we can deduce from Eq.(\ref{omegaeff}) the laser intensity $I(\Delta_{F'})$ as a function of the detuning $\Delta_{F'}$. \\

The rate of spontaneous emission for atoms starting in $\ket{F=1,m_F=-1}$ or $\ket{F=2,m_F=1}$ is:

\begin{equation} \label{R_sc_F}
R_\textrm{sc}^F = \Gamma \dfrac{I}{I_\textrm{sat}} \sum_{n=\left\{0,1\right\},F',\sigma=\pm}  \dfrac{ \left( M_{F,m_F}^{F',\sigma} \right)^2   }{1 +  4 \dfrac{\left( \Delta_{F,F'}^n \right)^2}{\Gamma^2}+\dfrac{I}{I_\textrm{sat}}} \mathcal{J}_n(\beta)^2
\end{equation}

\noindent where $\Delta_{F,F'}^n =   \omega_L + n \omega_\textrm{hfs} - \omega_F^{F'}$ is the detuning of the sideband $n$ (of frequency $\omega_L + n \omega_\textrm{hfs}$) relative to the transition $F \rightarrow F'$ (of energy $\hbar \omega_F^{F'}$). The ground state hyperfine splitting $\omega_\textrm{hfs}$ is also the EOM driving frequency. \\ 

From these, the probability of spontaneous emission for an entire interferometer can be calculated: 

\begin{equation} \label{ES totale}
P_\textrm{SE} = 1 - \exp \left(-R_\textrm{sc}^{1} \cdot 2 \tau -R_\textrm{sc}^{2} \cdot 2 \tau \right)
\end{equation}

\noindent where we consider that the atoms spend as much time ($2 \tau$) in state $\ket{F=1}$ and in state $\ket{F=2}$, the total Raman interaction duration being $4 \tau$. \\

We measure experimentally the number of atoms that are transferred from $\ket{F=1}$ to $\ket{F=2}$ when the Raman detuning is off the two-photon resonance. In order to compare these measurements with our theoretical model, we calculate the scattering rate $R_\textrm{sc}^{1 \rightarrow 2}$ for atoms initially in the state $\ket{F=1,m_F=-1}$, undergoing a single-photon transition to an excited state $\ket{F',m_{F'}}$, and transferred to $\ket{F=2}$ by spontaneous emission. This corresponds to the scattering rate $R_\textrm{sc}^1$  presented in Eq.(\ref{R_sc_F}) with the difference that one needs to take into account the spontaneous emission rates $\Gamma_{F' \rightarrow 2}$ of each de-excitation $\ket{F',m_{F'}} \rightarrow \ket{F=2,m_F}$. Therefore we have:

\begin{equation} \label{es f=2}
\begin{aligned}
R_\textrm{sc}^{1 \rightarrow 2} = \dfrac{I}{I_\textrm{sat}} \sum_{n=\left\{0,1\right\},F',\sigma=\pm} & \Gamma_{F' \rightarrow 2} \\
& \times \dfrac{(M_{1,-1}^{F',\sigma})^2}{1 +  4 \dfrac{(\Delta_{1,F'}^n)^2}{\Gamma^2}+\dfrac{I}{I_\textrm{sat}}} \mathcal{J}_n(\beta)^2
\end{aligned}
\end{equation}

\noindent where the spontaneous emission rates $\Gamma_{F' \rightarrow 2}$ are the following:

\begin{equation}
	\left\lbrace
\begin{aligned}
\Gamma_{0 \rightarrow 2} &=  0 \\
\Gamma_{1 \rightarrow 2} &=  \frac{\Gamma}{6}  \\
\Gamma_{2 \rightarrow 2} &=  \frac{\Gamma}{2}  \\
\Gamma_{3 \rightarrow 2} &=  0 \\
\end{aligned}
	\right.
\end{equation}

As presented in the associated article (see Figure 5), we introduce an empirical parameter $\kappa$ in the theoretical formulas to account for the difference between the effective Raman intensity and the theoretical prediction. Thus we write the total loss of atoms (\textit{i.e.} in the $\ket{F=1}$ and $\ket{F=2}$ levels) by spontaneous emission during the whole interferometer as :

\begin{equation} \label{ES totale avec k}
P_\textrm{SE} = 1 - \exp \left(-\kappa \left[ R_\textrm{sc}^{1} \cdot 2 \tau +R_\textrm{sc}^{2} \cdot 2 \tau \right]  \right)
\end{equation}

Likewise, we calculate the probability of transfer by spontaneous emission in $\ket{F=2}$ during a pulse of duration $\tau$ as follows :

\begin{equation} \label{ES F=2 avec k}
P_\textrm{SE~F=2} = 1 - \exp \left(-\kappa  R_\textrm{sc}^{1 \rightarrow 2} \cdot \tau\right)
\end{equation}

\section{One-photon and two-photon light shifts}

In a Mach-Zehnder type atom interferometer,  uncompensated differential light shifts $\delta_\textrm{LS}^\textrm{diff}$ from the Raman lasers result in an additional phase contribution given by \cite{Pereira2016}:

\begin{equation}
\begin{aligned}
\Delta \phi_\textrm{LS} =& -\arctan \left[\tan \left( \Omega_\textrm{eff,1} \dfrac{\tau}{2} \right) \dfrac{\delta_\textrm{LS,1}^\textrm{diff}}{\Omega_\textrm{eff,1}} \right]\\
& + \arctan \left[\tan \left( \Omega_\textrm{eff,3} \dfrac{\tau}{2} \right) \dfrac{\delta_\textrm{LS,3}^\textrm{diff}}{\Omega_\textrm{eff,3}} \right]
\end{aligned}
\end{equation}

\noindent where the effective Rabi frequencies and the differential light shifts for each pulse $i$  are given by $\Omega_\textrm{eff,i}$ and $\delta_\textrm{LS,i}^\textrm{diff}$ respectively. $\tau$ is the duration of the first and last pulses.  \\

The one-photon Raman light shift is imprinted onto both hyperfine states $\ket{F=1,m_F=-1}$ and $\ket{F=2,m_F=1}$ by out-of-resonance Raman lasers. Each light shift $\delta_\textrm{LS1}^{F,m_F}$ has the following expression:

\begin{equation}
\delta_\textrm{LS1}^{F,m_F} = \dfrac{\Gamma^2}{4} \dfrac{I}{I_\textrm{sat}} \sum_{n=\left\{0,1\right\},F',\sigma=\pm}  \dfrac{\left( M_{F,m_F}^{F',\sigma}\right)^2  }{\Delta_{F,F'}^n } \mathcal{J}_n(\beta)^2
\end{equation}

The differential one-photon light shift is then:

\begin{equation}
\delta_{\textrm{LS1}}^\textrm{diff} = \delta_\textrm{LS1}^{2,1}  -\delta_\textrm{LS1}^{1,-1} 
\end{equation}

There is no intensity ratio that cancels out the differential light shift, contrary to LPAIs using the $\ket{F=1,m_F=0} - \ket{F=2,m_F=0}$ transition. This means that a difference of intensity between the first and the third pulses leads to an additional phase shift \cite{Pereira2016}. However this light shift is nearly rejected through the $k$-reversal technique, since $\delta_{\textrm{LS1}}^\textrm{diff}(+k_\textrm{eff}) \simeq \delta_{\textrm{LS1}}^\textrm{diff} (-k_\textrm{eff})$. We recall that reversing the sign of $k_\textrm{eff}$ means preparing the atoms alternatively in the $\ket{F=1,m_F=\mp1}$ states. \\

In our setup, the two-photon light shift arises from the off-resonant copropagating Raman transitions detuned by $ \pm \Delta_B = \pm 2 \times \frac{\mu_B g_F}{h} B $  from the considered $\pm k_\textrm{eff}$ counterpropagating transition. The level $\ket{F=1,m_F=-1}$ is perturbed by the coupling $\Omega_\textrm{eff,co}^{-1}$ of the copropagating transition $\ket{F=1,m_F=-1} \leftrightarrow  \ket{F=2,m_F=-1}$. Likewise, the level $\ket{F=2,m_F=1}$ is perturbed by the coupling $\Omega_\textrm{eff,co}^{+1}$ of the copropagating transition $\ket{F=1,m_F=1} \leftrightarrow  \ket{F=2,m_F=1}$. The corresponding light shifts are:
 
\begin{equation}
	\left\lbrace
\begin{aligned}
\delta_\textrm{LS2}^{1,-1} &=  \dfrac{ \left(\Omega_\textrm{eff,co}^{-1} \right) ^2}{4 \Delta_B} \\
\delta_\textrm{LS2}^{2,1} &=  \dfrac{ \left(\Omega_\textrm{eff,co}^{+1} \right) ^2}{-4 \Delta_B} \\
\end{aligned}
	\right.
\end{equation}

\noindent where $\Delta_B =  2 \times \frac{\mu_B g_F}{h} B $ is the first order Zeeman splitting between the magnetic sublevels $m_F=0$ and $m_F=2$. The effective Rabi frequency of the copropagating transitions are:

\begin{equation} 
\left \lbrace
\begin{aligned}
\Omega_\textrm{eff,co}^{-1} &= \dfrac{\Gamma^2}{2}  \dfrac{I}{I_\textrm{sat}} \mathcal{J}_0(\beta) \mathcal{J}_1(\beta) \sum_{F',\sigma = \pm}  \dfrac{M_{1,-1}^{F',\sigma} \cdot M_{2,-1}^{F',\sigma}}{\Delta_{F'}} \\
\Omega_\textrm{eff,co}^{+1} &= \dfrac{\Gamma^2}{2}  \dfrac{I}{I_\textrm{sat}} \mathcal{J}_0(\beta) \mathcal{J}_1(\beta) \sum_{F',\sigma = \pm}  \dfrac{M_{1,1}^{F',\sigma} \cdot M_{2,1}^{F',\sigma}}{\Delta_{F'}}
\end{aligned}
\right.
\end{equation}

Finally, the differential two-photon light shift is:

\begin{equation}
\delta_{\textrm{LS2}}^\textrm{diff} = \delta_\textrm{LS2}^{2,1}  -\delta_\textrm{LS2}^{1,-1} 
\end{equation}

Just as for the one-photon light shift, the differential two-photon light shift is almost completely rejected through the $k$-reversal technique, because $\delta_{\textrm{LS2}}^\textrm{diff}(+k_\textrm{eff}) \simeq \delta_{\textrm{LS2}}^\textrm{diff} (-k_\textrm{eff})$.

\newpage
\bibliography{bibliography_of_article}

\end{document}